\documentclass{aa}  

\usepackage{natbib}
\usepackage{graphicx}
\usepackage{txfonts}
\usepackage{amsmath}
\usepackage{amsfonts}
\usepackage{url}

\newcommand{\mission}[1]{\textit{#1}}
\newcommand{\optprefix}[1]{(#1\mbox{-})}
\newcommand\blfootnote[1]{%
  \begingroup
  \renewcommand\thefootnote{}\footnote{#1}%
  \addtocounter{footnote}{-1}%
  \endgroup
}

\begin{document}

\title{Cyclotron resonant scattering feature simulations}

\subtitle{II. Description of the CRSF simulation process}

\author{F.-W.~Schwarm\inst{1},
	R.~Ballhausen\inst{1},
	S.~Falkner\inst{1},
	G.~Sch\"onherr\inst{2},
	K.~Pottschmidt\inst{3,4},
	M.~T.~Wolff\inst{5},
	P.~A.~Becker\inst{6},
	F.~F\"urst\inst{7},
	D.~M.~Marcu-Cheatham\inst{3,4},
	P.~B.~Hemphill\inst{8},
	E.~Sokolova-Lapa\inst{9,10},
	T.~Dauser\inst{1},
	D.~Klochkov\inst{11},
	C.~Ferrigno\inst{12},
	J.~Wilms\inst{1}
}
\authorrunning{F.-W. Schwarm et al.}

\institute{Dr.~Karl Remeis-Sternwarte and Erlangen Centre for Astroparticle Physics, Sternwartstrasse 7, 96049 Bamberg, Germany
\and Leibniz-Institut f\"ur Astrophysik Potsdam (AIP), An der Sternwarte 16, 14482 Potsdam, Germany
\and CRESST, Department of Physics, and Center for Space Science and Technology, UMBC, Baltimore, MD 21250, USA
\and NASA Goddard Space Flight Center, Code 661, Greenbelt, MD 20771, USA
\and Space Science Division, Naval Research Laboratory, Washington, DC 20375-5352, USA
\and Department of Physics \& Astronomy, George Mason University, Fairfax, VA 22030-4444, USA
\and European Space Astronomy Centre (ESA/ESAC), Science Operations Department, Villanueva de la Ca\~nada (Madrid), Spain
\and Kavli Institute for Astrophysics and Space Research, Massachusetts Institute of Technology, Cambridge, MA 02139, USA
\and Faculty of Physics, M. V. Lomonosov Moscow State University, Leninskie Gory, Moscow 119991, Russia
\and Sternberg Astronomical Institute, Moscow M. V. Lomonosov State University, Universitetskij pr., 13, Moscow 119992, Russia
\and Institut f\"ur Astronomie und Astrophysik, Universität Tübingen (IAAT), Sand 1, 72076 Tübingen, Germany
\and ISDC Data Center for Astrophysics, Université de Genève, chemin d’Écogia 16, 1290 Versoix, Switzerland
}

\date{Received December 14, 2016; accepted January 23, 2017}

\abstract
{
	Cyclotron resonant scattering features (CRSFs) are formed by scattering
of X-ray photons off quantized plasma electrons in the strong magnetic field
(of the order  $10^{12}$\,G) close to the surface of 
an accreting X-ray pulsar. Due to the complex scattering cross sections the line
profiles of CRSFs cannot be described by an analytic expression. Numerical
methods such as Monte Carlo (MC) simulations of the scattering processes
are required in order to predict precise line shapes for
a given physical setup, which can be compared to observations to gain information about the
underlying physics in these systems.
}
{
	A versatile simulation code is needed for the generation of synthetic
cyclotron lines. Sophisticated geometries
should be investigatable by making their simulation
possible for the first time.
}
{
	The simulation utilizes the mean free path tables described in
the first paper of this series for the fast interpolation of propagation
lengths. The code is parallelized to make the very time consuming simulations
possible on convenient time scales. Furthermore, it can generate responses to
mono-energetic photon injections, producing Green's functions,
which can be used later to generate spectra for arbitrary continua.
}
{
	We develop a new simulation code to generate synthetic cyclotron
lines
for
complex scenarios, allowing for unprecedented
physical interpretation of the observed data. An associated XSPEC 
model implementation is used to fit synthetic line profiles
to \mission{NuSTAR} data of \object{Cep X-4}.
The code has
been developed with the main goal of overcoming previous geometrical
constraints in MC simulations of CRSFs. By applying this code also to
more simple, classic geometries used in previous works, we furthermore address
issues of code verification and cross-comparison of various models. The
XSPEC model and the Green's function tables are available online
(see link in footnote, page 1).
}
{}

\maketitle

\section{Introduction}\label{sec:intro}
\blfootnote{\url{http://www.sternwarte.uni-erlangen.de/research/cyclo}}
Cyclotron resonant scattering features (CRSFs, or ``cyclotron lines'')
result from the interaction of photons with electrons in a strong
magnetic field ($B\gtrsim 10^{12}$\,G), for example, in the accretion
column of the magnetized neutron star in accreting X-ray binaries.
Their complex shape \citep[][and references
  therein]{isenberg98,schoenherr07,nishimura08,fuerst15,schwarm16a} reveals information about
the environment of the line forming region, which is typically close
to the neutron star.
A physical model for the line formation is needed in order to translate
observed line characteristics --- like the line's depth, width, and shape ---
into physically meaningful parameters. 

This paper is the second in a series in which we describe calculations of
resonant cyclotron scattering for various configurations of the
accreting matter. In the first paper of the series \citep[][paper~I in
  the following]{schwarm16a}, we described the calculation of the
relevant cross section for the range of conditions typically expected
for accreting neutron stars in X-ray binary systems. In this work we
discuss the application of these cross sections to the simulation of synthetic cyclotron lines.
Simulations of CRSFs have been performed by various groups in the
past. They can be divided into three classes: Monte Carlo
\citep[e.g.,][]{wang88,wang89,araya96,araya99,schoenherr07,nobili08a,nobili08b},
Feautrier \citep[e.g.,][]{meszaros85a,alexander89,nishimura08},
and \optprefix{semi}analytic methods \citep[e.g.,][]{wang93}. \citet{wang88} provide an
overview over the methods used in previous works. 
\citet{isenberg98} discuss the features of the fundamental approaches and
use multiple methods for the generation
of synthetic cyclotron lines in various parameter regimes.
For optical depths on the order of
$10^{-4}$--$10^{-3} \tau_\mathrm{T}$, where $\tau_\mathrm{T}$ is the Thomson optical
depth, Monte Carlo (MC) simulations are the suitable method
\citep{isenberg98}.
From all of the calculations mentioned above only \citet{schoenherr07}
provided a method for direct comparison between data and observations with
the corresponding large grids of parameters.

\citet{araya96} performed a first comparison of simulated MC line
profiles to data from the X-ray pulsar \object{A0535+26}. By convolving
precalculated Green's functions, describing the response of a CRSF
medium to monoenergetic photon injection, with arbitrary
continua, \citet{schoenherr07} disentangled the time consuming MC
simulations from the choice of continuum. This approach enabled
us to develop a code which can be used to compare MC simulated spectra
to observational data using standard X-ray astronomical data analysis
packages such as XSPEC \citep{arnaud96} or ISIS \citep{houck00}. A
disadvantage of these earlier calculations, however, is that they are
limited to very simple predefined geometries. In addition, while the
general properties of the line behavior are correct, unfortunately
there was an error in the integration of the scattering cross section
code by \citet[][see paper~I for details]{sina96}. 

Building on previous results and experience we have developed a new
simulation tool to calculate synthetic line profiles for arbitrarily
complex cylindrically symmetric geometries. This generalization naturally
requires considerable computational effort, which must be met with new
strategies. The mean free path (MFP) tables used for the interpolation
of the thermally averaged scattering cross sections have been
described in paper I. Here, we will provide the description and
applications of the full MC scattering code, which has been written
with the prime goal of imprinting cyclotron lines on the continuum
emission of accreting X-ray pulsars, and which includes a working
fit model. The XSPEC model code and the Green's tables necessary to
generate synthetic cyclotron lines are provided, as well as instructions
on how to use them. The discussion of more sophisticated physical
scenarios and the application to observational data will be the subject of successive papers in this series.
Here, we restrict ourselves to presenting examples of selected
synthetic spectra for illustration and for comparison to other works.

The structure of this paper is as follows. In Sect.~\ref{sec:simulation}
the physical assumptions are laid out and the simulation process is explained
in detail, including a description of the scattering process and commonly used
geometries. The duration of an exemplary simulation run is analyzed with
respect to the used number of CPUs and computer systems. This is followed by a
description of the Green's function scheme, which is utilized to enable
the fitting of the synthetic cyclotron lines to observational data, taking
a step toward the application of the simulation in Sect.~\ref{sec:application}.
In the same section, a comparison of synthetic CRSFs to previous works is performed.
Furthermore, the Green's functions are applied to a physical continuum
for several combinations of line parameters. Also, the availability of the
XSPEC model and the associated Green's function tables is discussed.
The XSPEC model is fitted to \mission{NuSTAR} data of \object{Cen X-4}
and the resulting magnetic field strength and cyclotron line shape
is compared to results from empirical models.
Additional technical details about the model usage and its parameters are
given in the Appendix.

\section{Simulation of cyclotron lines}\label{sec:simulation}
In this section we first describe the physical setup and explain our
Monte Carlo approach in detail before discussing how these simulations
can be sped up. We also introduce various classical accretion
geometries which we will be using in Sect.~\ref{sec:application} in
order to validate our results with respect to the ones from earlier works using
the MC method.

\subsection{Physical setup}
We consider the interactions of photons with electrons in strong
magnetic fields via the cyclotron scattering process. Other particles,
such as protons which might form ``proton cyclotron lines''
\citep{ibrahim02,ibrahim03} in the spectra at energies below the
electron cyclotron lines, are neglected. There are several other
processes altering the properties and paths of X-ray photons in the
vicinity of magnetic fields near the critical field strength \citep{canuto77},
\begin{equation}\label{eq:B_crit}
B_\mathrm{crit}=
\frac{m_\mathrm{e}^2 c^3}{e\hbar} = 4.413\times 10^{13}\,\mathrm{G}\,.
\end{equation}
Photon splitting, which describes the process of splitting one photon
into two photons, dominates over Compton scattering for very high
magnetic field strengths and densities \citep{adler71}. Pair
production becomes possible if the photon energy exceeds $2m_\mathrm{e}c^2$
\citep{daugherty83}. Apart from photon-electron interactions,
\citet{sina96} also analyzed the properties of M{\o}ller and Bhabha
scattering where electrons are scattered off electrons or positrons,
respectively. In the regime we are interested in, cyclotron scattering
is the dominant process and therefore we will neglect all the other
possible interactions.

As further discussed in paper~I, the momenta of the CRSF forming
electrons perpendicular to the direction of the $B$-field are
quantized to discrete values corresponding to the Landau energy
levels, \mbox{$E_n\sim n\cdot12\,\mathrm{keV} \cdot B/10^{12}$\,G}.
The electrons can move freely parallel to the field
lines. Inelastic scattering of photons off these electrons leads to
the formation of CRSFs which appear as absorption-like lines at the
fundamental Landau energy $E_1$ and its integer multiples \citep[][and
  references therein]{gnedin74,canuto77,araya96,araya97,schoenherr07}.

The cyclotron scattering cross section strongly depends on the
incident photon angle and energy \citep[paper~I
  and][]{canuto71,canuto77,ventura79,meszaros79}.
The angle $\vartheta_\mathrm{in}$ is measured with respect to the magnetic
field axis and mostly specified by its cosine, $\mu_\mathrm{in} = \cos \vartheta_\mathrm{in}$.
In the rest frame of
an electron, the photons' energies and propagation angles are Lorentz
boosted due to the electrons' motion parallel to the $B$-field. This
makes the accurate analytic treatment difficult \citep{canuto71}
because a photon's energy and angle are relativistically coupled in
the electron rest frame.

Cyclotron absorption and emission processes as well as resonant
scattering generate highly complex line profiles. In contrast to Compton
scattering which does not absorb or create photons, the number of
photons that contribute to the spectrum is not conserved
\citep{bonazzola79}. A photon can be absorbed by an electron if it has
the right energy and angle to excite the electron to a higher Landau
level. Almost immediately, the electron will emit ``spawned'' photons
during its successive de-excitation to the ground state. Since
de-excitation preferentially takes place to the next lower Landau
level \citep{latal86}, the majority of spawned photons will have
energies corresponding to the energy difference between neighboring
Landau levels. This is close to the fundamental energy $E_1$
regardless of the initial Landau level of the electron. The transition
process continues until the ground state is reached, effectively
increasing the number of photons within the medium.
Therefore complex computational
methods are necessary to determine the exact shape of cyclotron
resonance features.

\subsection{Monte Carlo simulation}\label{sec:mc}

The advantage of a Monte Carlo simulation is that its photon tracing approach
allows for the simulation of the radiative transfer in very complex setups of
the scattering medium. In general, in a MC simulation a seed photon is
generated at a place where the primary photons originate. The photon is
assigned an energy, position, and direction. A random number is then drawn from
an exponential distribution, which depends on the photon mean free path. This
random number is the optical depth that the photon will travel.  As discussed
in paper I, in order to speed this step up we use precalculated mean free path
tables (see Fig.~\ref{fig:flowchart} and Sect.~\ref{sec:flowchart} below for
more details).

The scattering geometry is realized in our simulation by describing the
medium with a list of cylinders with arbitrary dimensions, which may be combined
to model all kinds of cylindrically symmetric shapes of accretion columns. The
geometry of each cylinder is parameterized by its inner --- to allow for hollow
cylinders --- and outer radii and the distances of its bottom and top to the
neutron star surface. The physical properties inside are given by its
homogeneous density\footnote{The density is usually calculated from the optical
depth into a given direction and the corresponding extension of the cylinder.
For the classical ``\mbox{slab 1-0}'' and ``\mbox{slab 1-1}'' geometries the
direction parallel to the $B$-field axis is chosen, while in the ``cylinder''
geometry the optical depth is defined perpendicular to this axis (see
Fig.~\ref{fig:geometries})}, magnetic field, electron temperature, and velocity
towards the neutron star. Multiple cylinders can be stacked on top of, or
inside of each other in order to properly simulate parameter gradients.  Here,
we still use the simple geometries, which are explained in
Fig.~\ref{fig:geometries} together with their implied seed emission patterns,
for comparison of our results to \citet{isenberg98} in Sec.~\ref{sec:isenberg}.

The seed photons in a simulation run result from the
configured photon sources. Different emission patterns are available
for each individual source: a point source emits
photons isotropically from its static origin, a line source
corresponds to photon emission from a line along part of
the magnetic field axis, meaning that the height of each photon above
the hypothetical neutron star is sampled individually, a plane source
describes an emitting plane perpendicular to the $B$-field axis at a
given height. Variants emitting only photons upwards, that is, in the
direction of the $B$-field axis, or downwards are available for the
point and plane source types to provide a convenient way of preventing
unprocessed photons from showing up in the resulting data. 

The implementation of photon spawning is straightforward: successive photon
generation and propagation of the spawned photons from the coordinates of the
de-exiting electron to the point where the spawned photons are interacting with
or escaping the medium ensure self-consisted treatment.

The most time-intensive parts of processing are parallelized using
the \citet[MPI]{mpi94}, decreasing the required CPU time efficiently as shown in
Fig.~\ref{fig:mpi_speed}.
Because available computing power has increased significantly since
earlier approaches to this problem, we were also able to introduce
additional features to the simulation which allow us to specify the
conditions to be simulated in a more flexible way. For example, more
complex physical settings for the accretion column, including velocity
gradients, $B$-field gradients, and an inhomogeneous density
stratification.

\subsection{Description of the scattering process}\label{sec:flowchart}
\begin{figure}\centering
	\resizebox{0.95\hsize}{!}{\includegraphics{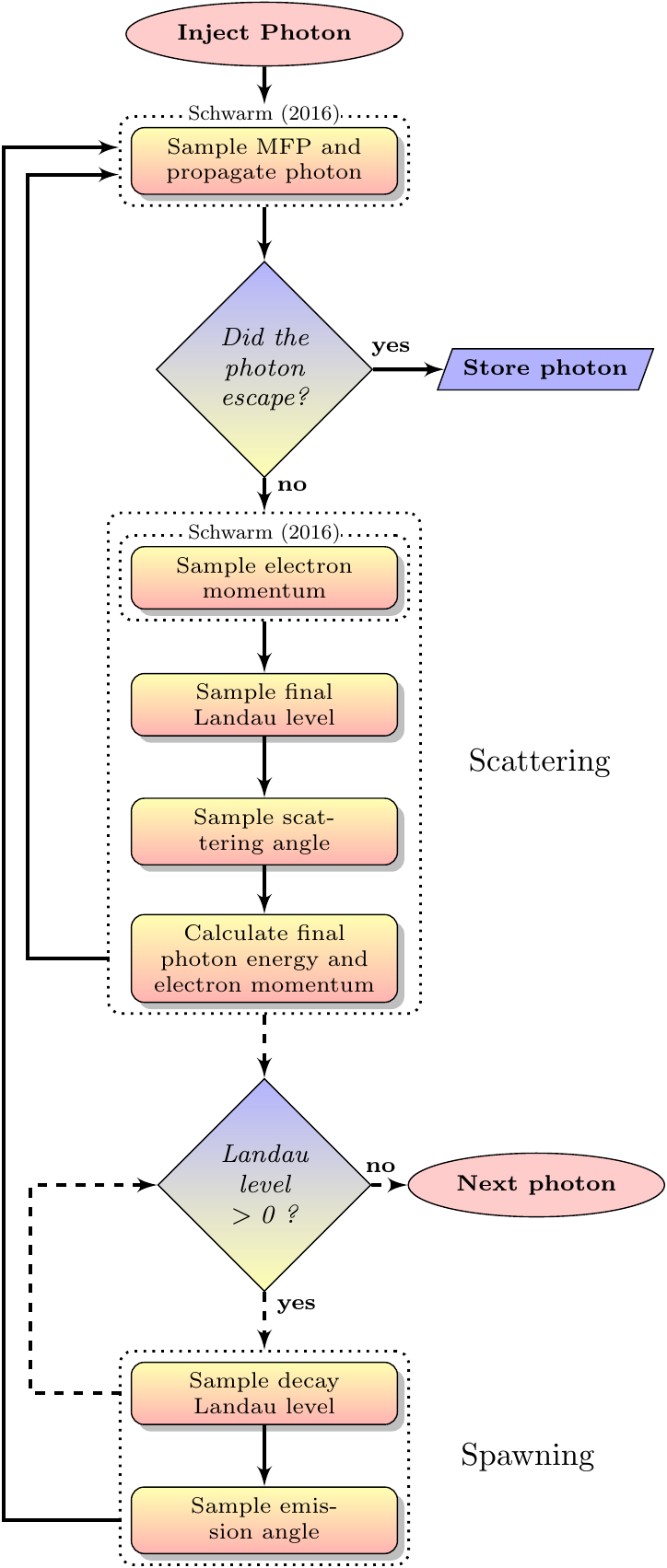}}
	\caption{Flowchart of the complete Monte Carlo process. The
          mean free path calculation and the electron momentum
          sampling block are described in detail in paper I. Solid lines
          correspond to photon related steps, dashed lines show
          electron related processes. Diamond shapes depict
          decisions.}\label{fig:flowchart}
\end{figure}
The core of the MC simulation is the propagation of individual photons
through a given CRSF medium, the process of which is illustrated by
Fig.~\ref{fig:flowchart}. Seed photons, generated by the configured
photon sources, must be processed as well as spawned photons,
originating from the transition of an electron to a lower Landau
level. For this purpose the seed photons are injected into the
simulation and photons spawned during their interactions are
added to the current list of photons and processed the same way.

Each photon is propagated according to its current path length, which
is sampled from the mean free path tables provided by
\citet{schwarm16a}. The photon is stored to a file in FITS format if
it escapes the medium and the simulation starts with the next seed
photon. Otherwise the momentum of the interacting electron is sampled,
again using precalculated tables (paper~I). We assume that the
electrons are in the ground state, that is, we neglect the
excitation of electrons through other processes than resonant cyclotron
scattering, which is justified by the comparably small time scales of
the electron decay compared to the collisional excitation rate \citep{bonazzola79,latal86,arons87,araya97,isenberg98}.
They can be excited to higher
Landau levels in the course of resonant cyclotron scattering, though. The final Landau level
for each interaction is sampled by comparing the scattering cross
sections for all possible final levels below the maximum Landau level
$n_{f,\mathrm{max}}$, which has been set to 5 for the simulations
performed here, in compliance with the number of Landau levels taken
into account for the MFP table calculations. In a similar manner, the
scattering angle is sampled by assigning a random number to the
cumulative angular distribution gained by integrating the cross
section over all possible final photon angles. The kinematic
calculations can be carried out once the initial properties of the
interacting electron and photon have been sampled, gaining the final
photon energy and electron momentum. The photon altered by the
scattering process is now further propagated, starting again with the
mean free path interpolation.

The excited electron will produce spawned photons during its
de-excitation to the ground state if it has been excited to a higher
Landau level during the scattering process. The number of spawned
photons depends on the Landau level the electron is excited to, which
is mainly determined by the photon energy in the electron rest frame.
De-excitation preferentially takes place to the next lower Landau
level, producing a photon with an energy corresponding to the energy
difference of the involved Landau levels. 
In order to show this quantitatively, we calculated the transition rates for a magnetic field of $0.1\,B_\mathrm{crit}$, averaged over initial and final spin and
summed over final polarization. They are shown in Table~\ref{table:decayrates} for initial Landau levels
up to $n_i = 7$. The sixths and seventh levels are provided as a reference and in order to allow
for estimations of the probability that a photon is spawned at the resonance energy of the maximum
Landau level taken into account in the simulation, for instance, via the transition $7 \rightarrow 2$.
The ratio between the rate for a transition from Landau level $n_i = 2$ to Landau level
$n_f = 1$, that is, $\overline{\Gamma}_{21} \approx 1.74 \times 10^{-3}$, and the corresponding
transition to the ground state, $\overline{\Gamma}_{20} \approx 2.44 \times 10^{-4}$, becomes
$\overline{\Gamma}_{21}/\overline{\Gamma}_{20} \approx 7.11$ in agreement with the
corresponding calculation by \citet[Table~5 therein]{latal86}. The ratio
is becoming smaller for larger fields as shown by \citet{latal86}, as well.

\begin{table*}
	\caption{Cyclotron emission rates for a magnetic field $B = 0.1\,B_\mathrm{crit}$,
averaged over initial and final electron spins and summed over all possible
polarization modes, in units of $\omega_\mathrm{B} = eB/m$ \citep[see, e.g.,][]{herold82}. The rates are given
for transitions from the initial Landau level $n_i \le 7$ (first column)
to all final Landau levels $n_f$ for which $n_f < n_i - 1$ (first row).
	}\label{table:decayrates}
	\centering
	\begin{tabular}{cccccccc}
\hline\hline
$n$& 0& 1& 2& 3& 4& 5& 6\\ 
\hline
1& $7.8708 \times 10^{-4}$ &  & & & & & \\ 
2& $2.4432 \times 10^{-4}$ & $1.7366 \times 10^{-3}$ & & & & & \\ 
3& $1.1313 \times 10^{-4}$ & $5.8499 \times 10^{-4}$ & $2.2486 \times 10^{-3}$ & & & & \\ 
4& $6.3951 \times 10^{-5}$ & $2.8415 \times 10^{-4}$ & $8.6338 \times 10^{-4}$ & $2.5362 \times 10^{-3}$ & & & \\ 
5& $4.0800 \times 10^{-5}$ & $1.6579 \times 10^{-4}$ & $4.4945 \times 10^{-4}$ & $1.0698 \times 10^{-3}$ & $2.6992 \times 10^{-3}$ & & \\ 
6& $2.8234 \times 10^{-5}$ & $1.0818 \times 10^{-4}$ & $2.7355 \times 10^{-4}$ & $5.8940 \times 10^{-4}$ & $1.2188 \times 10^{-3}$ & $2.7889 \times 10^{-3}$ & \\ 
7& $2.0710 \times 10^{-5}$ & $7.6123 \times 10^{-5}$ & $1.8357 \times 10^{-4}$ & $3.7232 \times 10^{-4}$ & $7.0200 \times 10^{-4}$ & $1.3252 \times 10^{-3}$ & $2.8336 \times 10^{-3}$ \\ 
\hline
	\end{tabular}
\end{table*}

Although the spacing of the
cyclotron resonances is not perfectly harmonic these photons have
energies very close to the fundamental energy in the electron's rest
frame. For Landau levels above the first excited state an electron
emits multiple photons during its successive de-excitation to the next
lower level until the ground state is reached. The spawned photons
have to be boosted to the neutron star frame of reference before they are further
processed, since the electron that re-emits a photon has some velocity
component parallel to the magnetic field due to the electrons' thermal momentum
distribution (see paper I).
The simulation code features the possibility to configure an additional velocity
component of the simulated medium. This leads to an additional boosting factor
for all photons entering the medium and can be used to simulate the bulk velocity
of the accreting matter. This has been
used by \citet{schoenherr14} for the calculation of phase dependent cyclotron
line spectra and emission patterns from a cylindrically symmetric accretion
column throughout which such a flow with relativistic velocities is expected
\citep[see, e.g.,][]{becker07}.
We neglect bulk motion in this work because it is unnecessary for the
purpose of code verification. For the application of the CRSF model to \object{Cep X-4}
the omission of bulk velocity is justified by the chosen geometry: in ``slab''
geometries the line forming region is assumed to reside at the neutron star hot
spot at which the matter has been decelerated to non-relativistic velocities.
The boosting caused by the thermal momentum distribution of the electrons, together with
the slightly anharmonic spacing of the Landau levels, is responsible
for the formation of line wings around the fundamental cyclotron line.
This apparent excess of photons occurs at energies slightly
above or below the first resonance $E_1$. The intermediate Landau
levels occupied by the electron during this process are sampled by
making use of the corresponding cyclotron decay rates. Each transition
produces a photon with an energy equal to the energy difference of the
Landau levels. Its angle is sampled according to the angle
dependent decay rates. All spawned photons are further propagated in
the same way as the seed photons until they leave the medium.

\subsection{Classic geometries}\label{sec:geo}
\begin{figure*}
\sidecaption
  \includegraphics[width=12cm]{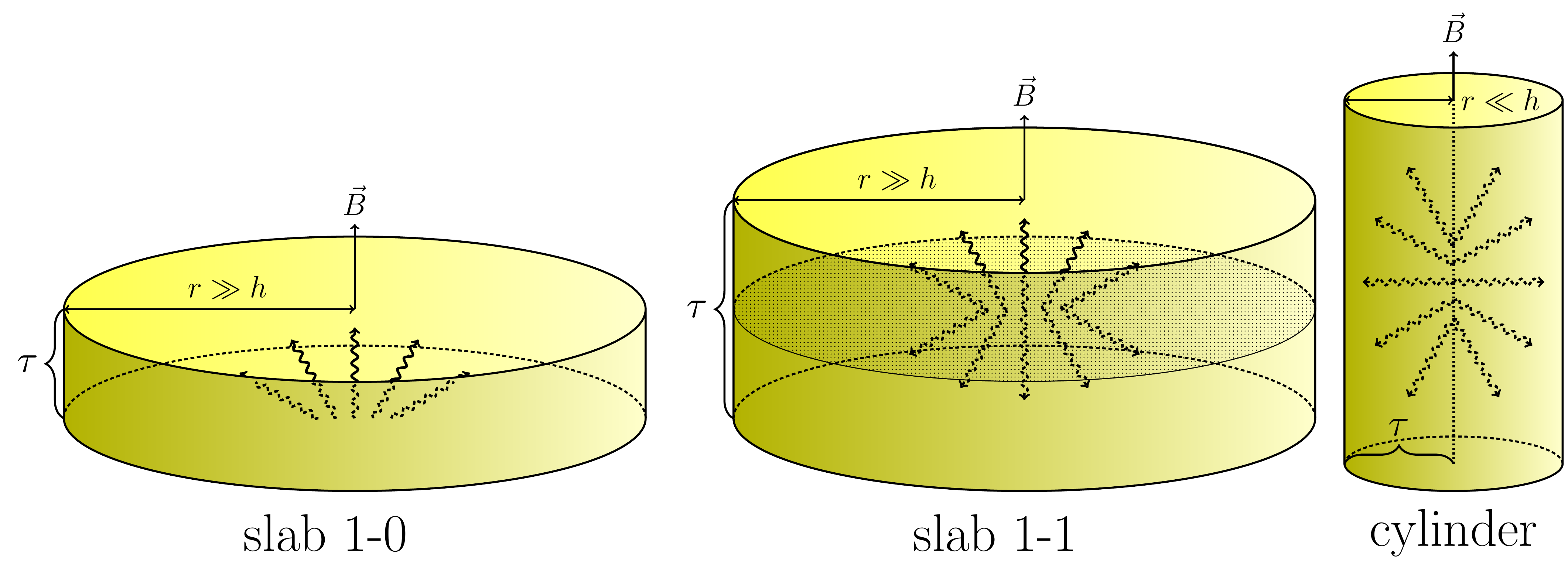}
  \caption{ Accretion column geometries used in earlier cyclotron
    calculations. The \mbox{slab 1-0} geometry is a bottom illuminated
    slab. The \mbox{slab 1-1} geometry corresponds to a slab
    illuminated by a seed photon plane in the middle of the slab. In
    cylinder geometry the seed photons are injected isotropically from
    a line in the center of the cylinder. Spectra emerging from the
    cylinder geometry are characterized by the optical depth in the
    radial direction, while the total optical depth along the vertical
    axis is used to define the scale of slab geometries.
  }\label{fig:geometries}
\end{figure*}

Figure~\ref{fig:geometries} depicts various geometries which differ in
radius to height ratio and the position, and type, of the seed photon source.

Slab geometries mimic a thin scattering volume of infinite radius. We
approximate this in our numerical simulations by setting the radius to a value
very large compared to the height of the medium. We find empirically that a
factor of 1000 between the radius and the height is a good choice in the
simulations. The output spectrum does not change significantly for a larger
extension in the radial direction, that is, only few photons, if any,
travel that far perpendicular to the $B$-field axis without escaping through
the bottom or top of the slab. The height of the slab is specified in terms of
the Thomson optical depth. The density in the slab can therefore be calculated in
the simulation from the slab height or radius, for \mbox{slab1-0}/\mbox{slab1-1}
or \mbox{cylinder} geometry, respectively, and the corresponding optical depth. For slab
geometries the optical depth parallel to the magnetic field is used to describe
the medium, because the slab radius is infinite or at least much larger than
its height. For a cylinder geometry the optical depth perpendicular to the
magnetic field axis is used.

Using the notation of \citet{isenberg98}, in the \mbox{slab 1-0}
geometry the medium is solely illuminated from the bottom, while in
the \mbox{slab 1-1} geometry a source plane in the middle produces seed
photons within the medium.

In the cylinder geometry the medium's radius is much smaller than its
height. Therefore, the optical depth perpendicular to the $B$-field
axis, that is, the cylinder radius in units of optical depths is used
as a parameter to describe the medium instead. Here, the seed photons
are also produced within the column but only along the cylinder's
axis.

In this work we show spectra for \mbox{slab 1-0} and \mbox{slab 1-1}
geometry in Fig.~\ref{fig:isenberg_comp} and spectra for cylinder
geometry in Fig.~\ref{fig:bw07}.
The CRSF shape is, in general, highly
sensitive to the simulated geometry. The formation of line wings, for
example, is especially pronounced in \mbox{slab 1-1} geometry for
viewing angles almost parallel to the magnetic field, which can be
seen in Fig.~\ref{fig:isenberg_comp}.

\subsection{CPU time}\label{sec:time}
\begin{figure}\centering
	\resizebox{\hsize}{!}{\includegraphics{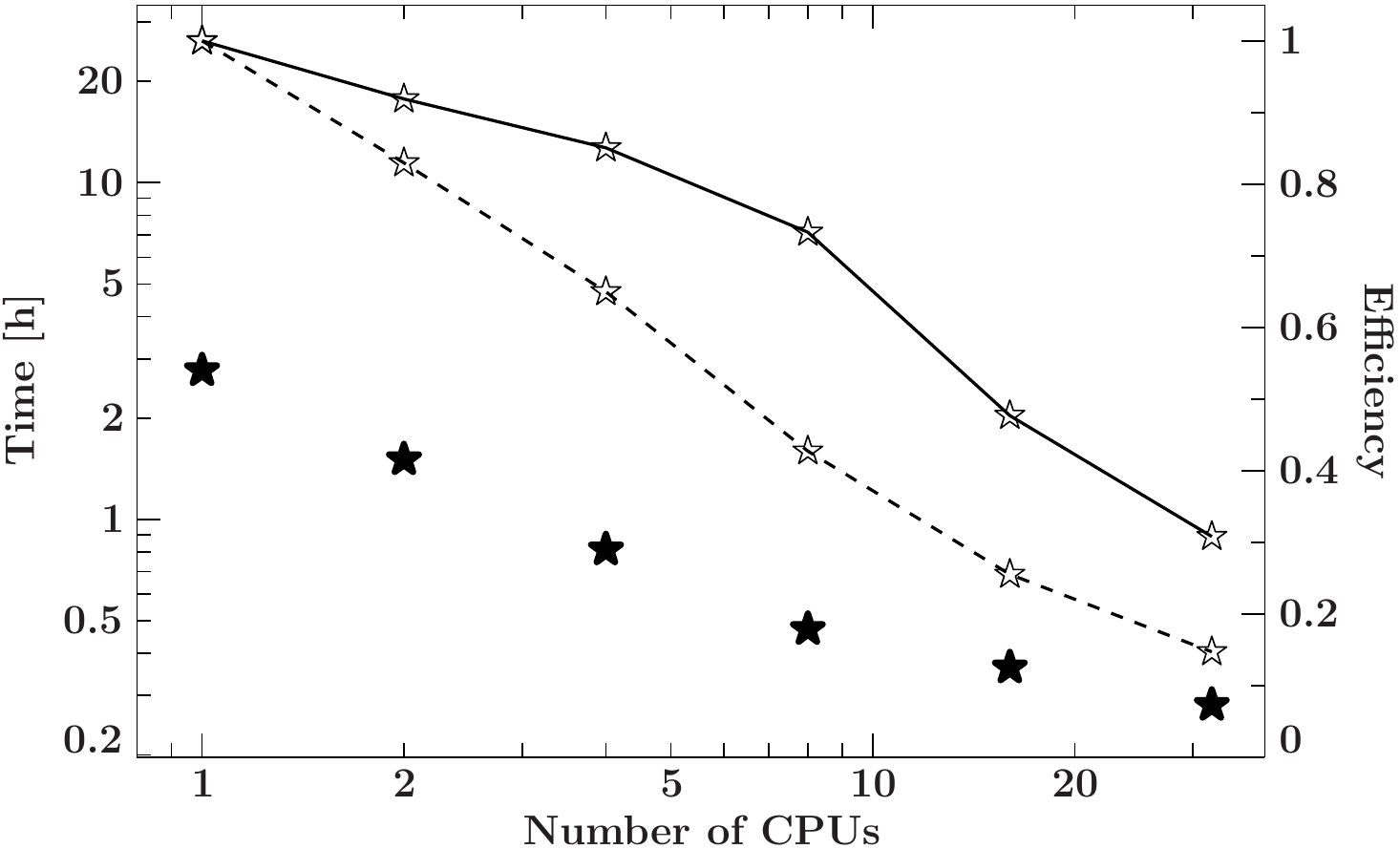}}
	\caption{Absolute CPU time (filled symbols) on the left axis
          and multi-core efficiency (lines) on the right axis
          for the simulation of synthetic spectra. The efficiency is
          calculated as $\eta = t_1 / (n t_n)$, where $t_{n}$ is the
          execution time on $n$ CPUs. The simulation speed tests have
          been performed on an AMD Opteron 2.2\,GHz system (dashed
          line) as well as on an Intel Xeon 2690 2.6\,GHz system (solid
          line and filled symbols). The simulations have been
          performed for the same parameters as in
          Fig.~\ref{fig:isenberg_comp} below, but with only 100 input
          photons per energy bin instead of 10000. Absolute CPU times
          depend on the chosen parameter
          values.}\label{fig:mpi_speed}
\end{figure}
We reduce the calculation time by using MPI to
distribute the work among multiple processors.
Figure~\ref{fig:mpi_speed} shows that our simulation is parallelized
efficiently. We performed tests on different computer systems and
found that using $\sim$8 CPUs for each simulation provides a
convenient compromise between simulation time and efficiency.

\subsection{Green's table approach}
Performing a time consuming simulation for each generation of a
spectrum is not practicable in anticipation of our goal to fit
simulated spectra to observational data. A large number of simulation
runs will be required where the shape of the emerging spectrum is
simulated as a function of the input parameters. In addition to the
magnetic field strength and the plasma temperature, the spectral shape
of the incident photons is also variable. This significantly increases
the computing time as potentially a large parameter space needs to be
covered. As it was shown previously \citep{schoenherr07},
this large expense of computing time can be avoided by calculating the
Green's function of the radiative transfer problem. We propagate
photons through a medium with a given geometry, magnetic field
strength, and temperature for a well sampled grid of mono-energetic
seed photon energies. The photons escaping the medium are collected
and binned to spectra for a grid of output angles. These emerging
spectra, normalized to the seed photon flux in a given
angular bin, correspond to the Green's function of the radiative
transfer problem. The emerging spectrum for an arbitrary seed photon
spectrum, like a cutoff power law continuum or a continuum due to
Comptonization in a radiative shock \citep{becker07}, can then be
obtained by convolving it with the Green's functions.

In general, our approach requires to interpolate the Green's
functions for the energies on which the seed photon continuum is
defined. This step is necessary because these energies are typically
defined by the response matrix of the detector with which the data
were taken. This energy grid is outside of our control. We have
calculated grids with a logarithmic energy spacing that is fine enough
that linear interpolation is sufficient to re-grid the Green's
functions in energy.

A second interpolation step in the magnetic field strength is required
to obtain a Green's function for a $B$-field value not covered by the
grid of precalculated values. Since the Green's functions are
self-similar in $E/B$ we interpolate the Green's functions for
different $B$ in this ``energy shifted'' system. In a similar way
further interpolations in temperature, output angle, and optical depth
are used to approximate the final spectrum for parameter combinations
off the grid points.

Using MFP interpolation tables, we are able to produce Green's
functions for a large parameter range and with a resolution better
than the resolution of all currently flying X-ray telescopes for the
energy range where cyclotron lines are observed. This takes about
300\,CPU hours for one geometry on rather coarse grained magnetic
field and temperature grids. In principle the MFP table mechanism, and
therefore the creation of Green's tables, is reasonable for magnetic
fields $0.01 \lesssim B/B_\mathrm{crit} \lesssim 0.12$ and
temperatures $0\,\mathrm{keV} \le T \lesssim 50\,\mathrm{keV}$. For
much smaller $B$-fields the fully relativistic quantum-electro-dynamic
calculations are unnecessarily complex, while for much higher fields
cyclotron scattering may not be the dominant process \citep[see,
  e.g.,][]{sina96}. Much higher
temperatures are not expected to occur in accreting X-ray
binaries. For the testing of different geometries and other parameters
we currently concentrate on an optimized subset in order to save CPU
time. Table~\ref{table:greens_list} lists the parameters we have
calculated Green's function tables for. The Green's functions needed
for the evaluation of model spectra for arbitrary parameters within
the precalculated ranges are interpolated from these tables using the
methods given in Appendix~\ref{sec:model_interp}. The Appendix also
describes the extrapolation beyond the covered parameter regime.
However, this extrapolation should be taken with extreme caution.

\section{Application}\label{sec:application}

We now compare
our results to those of previous works and discuss the generation of
synthetic CRSF spectra for a physical continuum. We use end-to-end
comparison to previous works since we do not have access to their
intermediate data products such as mean free path or momentum sampling
information. A comparison to a \mission{NuSTAR} observation of \object{Cep X-4}
is performed to show the applicability of the model to real data and
as a demonstration of the new CRSF fitting model.

\subsection{Code verification}\label{sec:isenberg}
\begin{figure}
	\resizebox{\hsize}{!}{\includegraphics{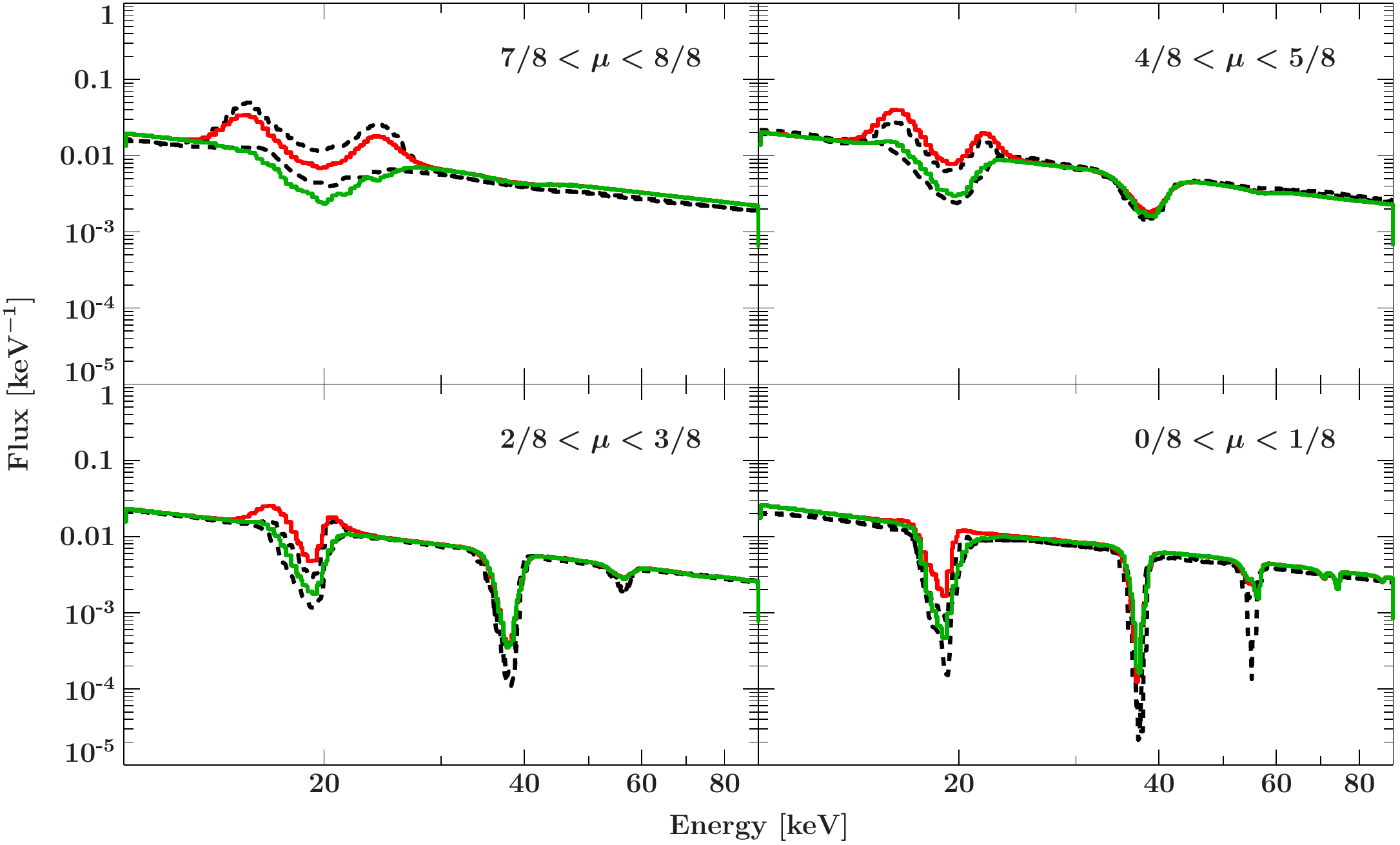}}
	\caption{Comparison of this work (solid lines) to \citet[their
            Fig.~7; dashed lines]{isenberg98}.
	Red lines denote the \mbox{slab 1-1} geometry, green lines the
          \mbox{slab 1-0} geometry.
The emission in different angular bins is labeled by
the cosine of the viewing angle to the magnetic field axis, $\mu = \cos \vartheta$.
Temperatures of 7\,keV and 8\,keV
          have been used for the \mbox{slab 1-0} and \mbox{slab 1-1}
          geometries, respectively.}\label{fig:isenberg_comp}
\end{figure}

In order to validate the general simulation,
Fig.~\ref{fig:isenberg_comp} compares spectra obtained using our
simulation with earlier calculations by \citet{isenberg98} for two
different accretion column geometries. Four different angular
regimes are shown. The upper left plot shows the simulated
spectra for a viewing angle almost parallel to the magnetic field
axis. The lower right plot shows spectra for viewing angles
almost perpendicular to the $B$-field axis. The corresponding angle
bins are defined in terms of the cosine of the viewing angle
$\mu = \cos \vartheta$.

The calculations agree well: the line width decreases with increasing
viewing angle, while the depth of the line decreases with decreasing viewing
angle. Indications for a third harmonic line can be seen in spectra
emerging perpendicular to the magnetic field, while for viewing angles
parallel to the magnetic field only a complex fundamental line is
observed. The strong line wings become less pronounced for
continua with a high energy cutoff, as less photons are spawned from
the higher harmonics \citep[][and references
  therein]{isenberg98,araya97,schoenherr07}:

Previous works based on the numerically calculated cyclotron cross
sections obtained by \citet{sina96}, such as \citet{araya97} and
\citet{schoenherr07}, do not show such a good agreement with the
spectra from \citet{isenberg98} due to their usage of erroneous
integrated cross sections in the simulation code \citep[][and
  paper~I]{schwarm12}.

Some slight deviations remain in the line wings. These are formed by
spawned photons. In the spectra of \citet{isenberg98} a larger number
of spawned photons escape at small angles to the magnetic field. These
extra photons appear at larger angles in our simulation. The reason
for this disagreement is probably a different angular redistribution
scheme or the approximation of the total
scattering cross section used by \citet[][Eq.~11 therein]{isenberg98}
for modeling the resonant scattering at harmonics above the first one.
\citet{isenberg98} use this approximation \citep[see
  also][]{daugherty77,fenimore88}, which is strictly valid only near
the line center, for resonant scattering involving the excitations of
higher harmonics. They resort to a numerical integration over the
scattering angle of the more correct form given in their Eq.~10
\citep[see
  also][]{canuto71,herold79,ventura79,wasserman80,harding91,graziani93}
for transitions involving only the ground state and the first excited
state.

Although the discussion of cyclotron resonant scattering as a cooling
process for the electrons in the accretion column is beyond the scope
of this paper, we briefly give reasoning for the two different
temperatures used in the comparison. The electron temperature depends
on the photon interactions which in turn depend on the electron
temperature. In the regime where cyclotron resonant scattering is the
dominant cooling process, a convenient choice for an equilibrium
temperature is the temperature at which the energy transfer is zero.
The point of equilibrium depends on the geometry.
Figure~\ref{fig:isenberg_comp} therefore assumes temperatures of
7\,keV and 8\,keV for the \mbox{slab 1-0} and \mbox{slab 1-1}
geometries, respectively, following \citet{isenberg98}. These
temperatures correspond to the Compton temperature where the energy
transfer is minimal. \citet{lamb90} used this technique to show that
the ratio of this equilibrium temperature to the magnetic field is
fairly constant if the temperature is determined solely due to
cyclotron resonant scattering.

\subsection{Cyclotron lines for the \citet{becker07} continuum}\label{sec:bw07}
\begin{figure}
	\resizebox{\hsize}{!}{\includegraphics{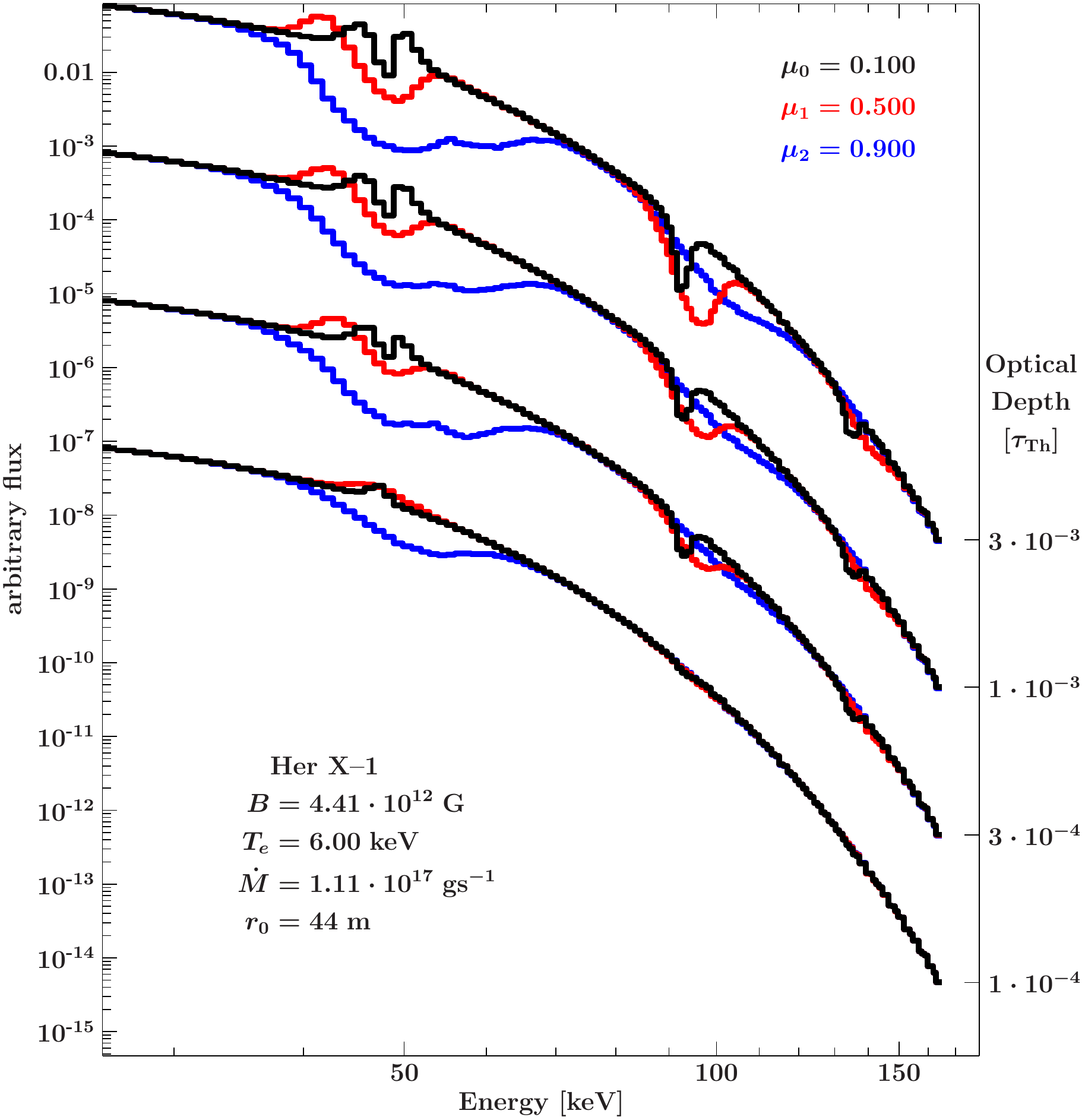}}
	\caption{Continuum from \citet{becker07} with imprinted
          cyclotron lines for several angles $\mu = \cos \vartheta$
	to the magnetic field axis and optical depths $\tau$. The
          continuum parameters are the same as in the
          spectrum from \citet[Fig.~6 therein]{becker07} for
          \object{Her X-1}. The different colors show three angles to
          the magnetic field. The optical depth is increasing from the
          bottom to the top as shown by the axis on the right side.}
	\label{fig:bw07}
\end{figure}

Figure~\ref{fig:bw07} shows synthetic cyclotron lines imprinted on the
continuum from \citet{becker07} for the purpose of illustrating
the influence of the continuum on the CRSFs and as an example
for the cylinder geometry.
The continuum parameters are the same ones as in the theoretical calculation
from \citet{becker07} for \object{Her X-1}. This calculation agrees very well
with the \mission{BeppoSAX} data reported by \citet{dalfiume98}. A spectral fit
of a recent XSPEC implementation of the same continuum model to
\mission{NuSTAR} data of \object{Her X-1} can be found in the work by
\citet{wolff16}.
Figure~\ref{fig:bw07} shows that the lines become deeper with increasing
optical depth. The line width decreases with increasing viewing angle
to the magnetic field, because of the smaller influence of Lorentz
boosting in this regime. A cylinder geometry Green's table has been
used for imprinting the cyclotron lines on the continuum. This
approach leads to emission like behavior for very small optical depths
and large angles to the magnetic field. Two higher harmonics can
clearly be seen and are especially pronounced for large angles and
relatively high optical depths of $\tau = 3 \cdot
10^{-3}\,\tau_\mathrm{T}$.

These are theoretical spectra, which in contrast to observed spectra
correspond to a line forming region with a constant magnetic field
seen from a specific angle. Observations are expected to be smeared
out due to an angle mixing with phase and an extended line forming
region with a magnetic field gradient. Though the accurate handling of
angle mixing requires the inclusion of relativistic effects
\citep{schoenherr14}, its influence on the CRSF line profiles, in the
reference frame of the neutron star, can be roughly estimated by averaging over
multiple viewing angles. Appendix~\ref{sec:model_mu} provides details
on how this can be done easily with the model presented in the
following. All continua used in this work are angle averaged
but the combination of the model with angle dependent continua is
straight forward.

\subsection{The CRSF model and table availability}\label{sec:tables}

Together with this work we distribute improved Green's function tables
for the classical geometries, which can be used together with an XSPEC
local model, \texttt{cyclofs}, to imprint cyclotron lines on arbitrary
continua. See Appendix~\ref{sec:model} for a description of the
model. Table~\ref{table:greens_list} shows the parameter ranges
covered by these tables. Each table corresponds to one geometry. The
tables have been calculated using thermally and polarization averaged
cross sections.

The \texttt{cyclofs} model convolves the Green's table set by the user with the
given continuum. It can, in principle, extrapolate beyond the
ranges provided in Table~\ref{table:greens_list}, but this should be
used with care and currently triggers a warning message. The model and
preliminary Green's function tables are available online (see link in
footnote, page 1). The currently rather coarse grained parameter grids
will be refined successively and will be made available at the same
location.

\begin{table}
	\caption{Parameter ranges covered by CRSF Green's function tables}\label{table:greens_list}
	\centering
	\begin{tabular}{ccc}
\hline\hline
Parameter                                  & Minimum           & Maximum           \\
\hline
Magnetic field [$B_{\mathrm{crit}}$]       & 0.01              & 0.12              \\
Electron temperature $k_\mathrm{B}T$ [keV] & \phantom{0}3      & 15                \\
Cosine to $B$-field, $\mu = \cos\vartheta$   & 0.001             & 0.999             \\
Optical depth [$\tau_\mathrm{T}$]          & $1 \cdot 10^{-4}$ & $3 \cdot 10^{-3}$ \\
\hline
	\end{tabular}
\end{table}

\subsection{Fitting \object{Cep X-4}}\label{sec:cepx4}

In order to demonstrate the applicability of the model to observed data, a
comparison between empirically fitted line profiles with synthetic ones from
the simulation described above will be performed in the following. The results
further motivate the necessity of a cyclotron line model based on firm physical
grounds for the description of CRSF line profiles.

\object{Cep X-4}, also known as \object{GS 2138+56}, was discovered in
1972 June and July observations with the \mission{OSO-7} X-ray
telescope \citep{ulmer73}. It is an accreting high mass X-ray binary
(HMXB) with a pulse period of $\sim66\,\mathrm{s}$ as found
in its 1988 outburst with \mission{Ginga}
\citep{makino88a,makino88b} and later confirmed by \citet{koyama91}. The optical
counter part is a Be star \citep{roche97} at a distance of $3.8 \pm
0.6\,\mathrm{kpc}$ \citep{bonnet98}. \citet{koyama91} found that the
addition of a cyclotron line at $31\,\mathrm{keV}$ improved the fit
but they did not include it in their discussion for the sake of
comparison to other cataloged X-ray pulsars. \citet{mihara91} proposed
a cyclotron line at an energy of $30.5\,\pm\,0.4\,\mathrm{keV}$ and
deduced a magnetic field of $2.6 \times 10^{12}\,(1+z)\,\mathrm{G}$
for the source.
\mission{RXTE}/PCA observed further outbursts in 1997 and 2002.
\citet{mcbride07} performed a spectral and timing analysis of the
latter outburst and found a cyclotron line at
$30.7^{+1.8}_{-1.9}\,\mathrm{keV}$. \mission{NuSTAR}
\citep{harrison13} observed \object{Cep X-4} close to the maximum of
the outburst and during its decline on 2014 June 18/19 and 2014 July
1/2, respectively \citep{fuerst15}. The cyclotron line was measured at
an energy of $\sim30$~keV in both observations. \citet{fuerst15}
found that its shape deviates from a simple Gaussian line profile \citep[see also][]{jaisawal15},
which makes this observation a good candidate for an application of
the physical CRSF model described in this work.

We re-extracted the \mission{NuSTAR} data from ObsID 80002016002 (2014
June 18/19, exposure time 40.5\,ks) near the maximum of the outburst
using the same settings and
procedure as described by \citet{fuerst15} but using the CalDB version
20160922 and the \mission{NuSTAR} data analysis software (NuSTARDAS)
version 1.6.0 as distributed with HEASOFT 6.19. The source and
background spectra for focal plane module A and B (FPMA and FPMB) were
extracted separately, using circular regions with radii of 120\arcsec.
We use data in the 3.6--55\,keV band and rebinned the data to a
signal-to-noise (S/N) of 10 below 45\,keV and a S/N of 5 above that.

We use the same empirical continuum model as \citet{fuerst15},
which was already found by \citet{mcbride07} to describe the continuum
well, that is, a power-law with a Fermi-Dirac cutoff \citep{tanaka86,makishima92},
\begin{equation}\label{eq:fermidirac}
	F(E) = A E^{-\Gamma} \frac {1} {1 + e^{-(E - E_\mathrm{cut}) / E_\mathrm{fold}}}\,,
\end{equation}
with normalization constant $A$, photon index $\Gamma$, cutoff energy $E_\mathrm{cut}$,
and folding energy $E_\mathrm{fold}$.

An improved version of the \texttt{tbabs} model, namely \texttt{tbnew}\footnote{
\url{http://pulsar.sternwarte.uni-erlangen.de/wilms/research/tbabs/}}, is used
with abundances by \citet{wilms00} and cross sections by \citet{verner96} in
order to account for photoelectric absorption of the continuum.  A narrow iron
K$\alpha$ line has been used as in the analysis by \citet{fuerst15}. Their low
energy black body was dismissed because it did not improve the fit. Also,
contrary to the analysis by \citeauthor{fuerst15}, but in agreement with the
continuum model used by \citet{mcbride07}, we added a ``10 keV'' feature, that
is, a Gaussian emission line that facilitates the modeling of the continuum in
the $\sim$8--20\,keV range, which has been applied before, for various sources
and instruments \citep{mihara95,coburn02}. We found the width of this broad
emission component to be in agreement with but much better constrained than the
corresponding component used by \citet{mcbride07}. The centroid energy of $16.1
/ 1.3 = 12.4\,\mathrm{keV}$ is below the value of 14.4\,keV found by
\citet{mcbride07}. The additional factor of 1.3 is necessary for the comparison
because of the gravitational redshift: in contrast to previous modeling
approaches, our model includes a redshifting of all components but the iron
line with a fixed redshift of $z = 0.3$ \citep[see, e.g.,][]{schoenherr07}. The
XSPEC convolution model \texttt{zashift} is used for that purpose.

The luminosity of 1--$6 \times 10^{36}\,\mathrm{erg}\,\mathrm{s}^{-1}$ \citep{fuerst15}
is low compared to other accreting X-ray pulsars exhibiting cyclotron
lines \citep{fuerst15} and well below the theoretical limit where a
shock in the accretion column is expected to form \citep{becker12}.
This suggests the usage of a slab geometry for the CRSF model,
\texttt{cyclofs}, which is the only model that we use to fit the CRSF
line profile, meaning that we do not use any other absorption line
component.

The complete fit model used is thus:
\texttt{C$_\mathrm{FPMA}$ $\times$ tbabs $\times$
	(redshift(cyclofs(powerlaw $\times$ fdcut + gauss$_\mathrm{10\,keV}$)) +
	gauss$_\mathrm{iron}$)},
where \texttt{C$_\mathrm{FPMA}$} is the cross-calibration of
focal plane module FPMB relative to FPMA.

\begin{figure}
	\centering
	\resizebox{0.95\hsize}{!}{\includegraphics{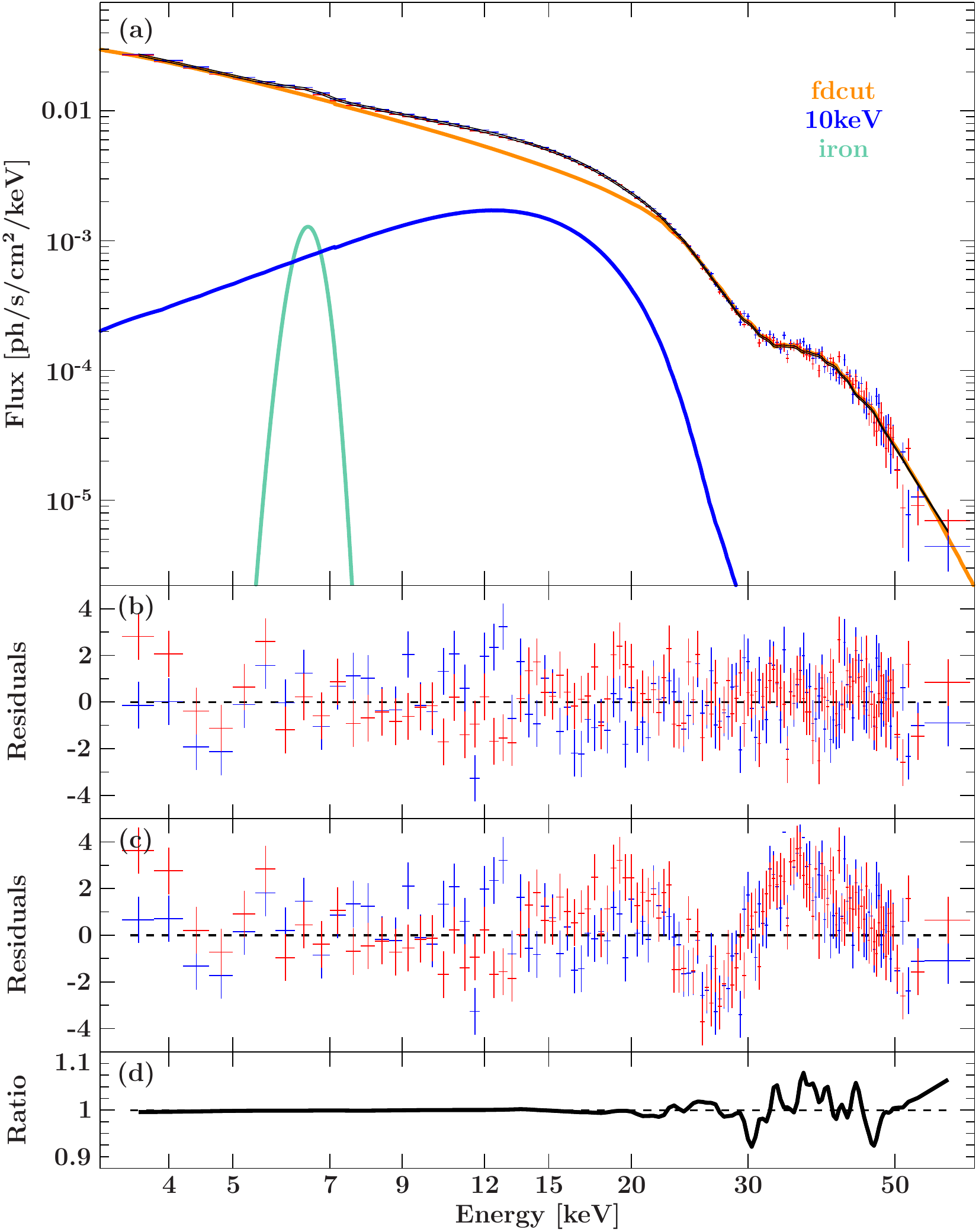}}
	\caption{(a) Unfolded spectrum, best-fit model, and individual
          continuum components for an observation (ObsID 80002016002)
          of Cep X-4. The data and residuals from \mission{NuSTAR}
          FPMA and FPMB are shown in red and blue, respectively. (b)
          Residuals to the best-fit model. (c) Residuals to the
          best-fit model with a Gaussian absorption line centered at
          the cyclotron line position found by the best-fit model.
          Only the width and depth of the line have been fitted. (d)
          Ratio between the physical and the empirical model. In the
          empirical model, all line parameters were allowed to vary
          including the centroid energy. The data have been regrouped
          for improved clarity.}
	\label{fig:fit}
\end{figure}

Starting with the parameters obtained from the fits by
\citet{mcbride07} and \citet{fuerst15}, the physical CRSF model for a
slab 1-0 geometry
converged towards an acceptable fit with a reduced $\chi^2$ of 1.17
for 862 degrees of freedom.
The unfolded spectrum is shown in Fig.~\ref{fig:fit}a together with
the model for both detectors, FPMA and FPMB, and the corresponding
residuals are shown in Fig.~\ref{fig:fit}b. The best fit parameters can be found in
Table~\ref{tbl:fit}, with uncertainties given at the 90\% confidence
level.

\begin{table}
	\caption{Best-fit parameters for the \texttt{cyclofs} model in
combination with a Fermi-Dirac cutoff continuum model.}
	\label{tbl:fit}
	\centering
	\begin{tabular}[b]{ccc}
\hline
Parameter & Value & Unit \\
\hline
$C_\mathrm{FPMB}$                    & $1.0341\pm0.0020$                                &                                           \\
$N_H$                                & $2.1^{+0.6}_{-0.5}$                              & $10^{22}$ cm$^{-2}$                       \\
$A_\mathrm{cont}$\tablefootmark{a}   & $0.230^{+0.036}_{-0.025}$                        & \\
$\Gamma$                             & $1.30^{+0.09}_{-0.07}$                           &                                           \\
$E_\mathrm{cut}$                     & $29.1^{+2.2}_{-1.9}$                             & keV                                       \\
$E_\mathrm{fold}$                    & $9.8^{+0.6}_{-0.4}$                              & keV                                       \\
$A_\mathrm{10 keV}$\tablefootmark{a} & $0.025^{+0.009}_{-0.006}$                        & \\
$E_{10\,\mathrm{keV}}$               & $16.1^{+0.5}_{-0.7}$                             & keV                                       \\
$\sigma_{10\,\mathrm{keV}}$          & $6.0^{+0.7}_{-0.6}$                              & keV                                       \\
$\tau_\parallel$                     & $\left(2.59^{+0.16}_{-0.24}\right)\times10^{-4}$ & $\tau_\mathrm{Th}$                        \\
B                                    & $3.627^{+0.030}_{-0.031}$                        & $10^{12}\,\mathrm{G}$                                  \\
T\tablefootmark{b}                   & $14.4^{+0.5}_{-0.9}$                             & keV                                       \\
$\mu = \cos \vartheta$               & $0.82^{+0.04}_{-0.05}$                           &                                           \\
$A_\mathrm{iron}$\tablefootmark{a}   & $\left(9.6^{+1.2}_{-1.0}\right)\times10^{-4}$    & \\
$E_\mathrm{iron}$                    & $6.491^{+0.027}_{-0.028}$                        & keV                                       \\
$\sigma_\mathrm{iron}$               & $0.30\pm0.05$                                    & keV                                       \\
red. $\chi^2$                        & 1.17                                             &                                           \\
d.o.f.                               & 862                                              &                                           \\
\hline
	\end{tabular}
	\tablefoot{\\
		\tablefoottext{a}{In photons keV$^{-1} $s$^{-1} $cm$^{-2}$ at 1 keV}. \\
		\tablefoottext{b}{The electron temperature in this fit is only
slightly below the current upper boundary of $15\,\mathrm{keV}$ covered by the
table. The corresponding upper error is therefore given at the $68\%$
confidence level.}
	}
\end{table}

As discussed by \citet{mueller13}, the position of the cyclotron line
can be affected by the continuum model. This complicates the
comparison of the magnetic field strength to previous works.
\citet{koyama91} noticed an absorption feature at $\sim
31\,\mathrm{keV}$ using a power-law plus exponential cutoff.
\citet{mihara91} used a power-law times cyclotron scattering cutoff
\citep{tanaka86} to describe the continuum and found the feature to
reside at $30.5 \pm 0.4\,\mathrm{keV}$. Using a model consisting of
negative and positive power-laws with a common exponential cutoff
\citep[NPEX]{mihara95}, \citet{makishima99} detected the cyclotron
line at $28.8 \pm 0.4 \,\mathrm{keV}$.
\citet{jaisawal15} included a fundamental cyclotron line at $27.5 \pm 0.4\,\mathrm{keV}$,
$27.7 \pm 0.4\,\mathrm{keV}$, or $29.6 \pm 0.5\,\mathrm{keV}$ in their
best-fit models for an NPEX, CompTT, or Fermi-Dirac plus blackbody continuum model,
respectively, in their analysis of the \mission{Suzaku} observation of \object{Cep X-4}
in 2014. The corresponding widths of the Gaussian absorption
lines used for fitting the CRSFs differ from each other as well between
the different continuum models.
\citet{mcbride07} and
\citet{fuerst15}, both using a Fermi-Dirac continuum model, found cyclotron lines at
$30.7_{-1.9}^{+1.8}\,\mathrm{keV}$ and
$30.39_{-0.14}^{+0.17}\,\mathrm{keV}$ (and
$29.42_{-0.24}^{+0.27}\,\mathrm{keV}$ during the decline of the 2014
outburst), respectively.
The analyses of all these works differ by more than the continuum model: some
use additional \texttt{gabs} model components to model the continuum and/or asymmetries in
the fundamental line profile, different models are used to describe the line
shape including Gaussian absorption lines and pseudo-Lorentzian profiles such
as \texttt{cyclabs}, different instruments might be responsible for systematic
deviations, and the physical behaviour of the source itself, such as variations
of the height of the line forming region with luminosity \citep[see,
e.g.,][]{becker12}, might lead to different magnetic field strengths --- and
therefore differences of the measured cyclotron line energy --- between
observations. The range of the values from the previous works listed above,
from $\sim28\,\mathrm{keV}$ to $\sim31\,\mathrm{keV}$, and differences of
$\sim8\%$ resulting from the application of different continuum models to
the same observation further illustrate the incomparableness of
cyclotron line energies resulting from differing analyses.

In order to compare the physical line model with an empirical model of the line
shape on the basis of the same continuum, we replaced the \texttt{cyclofs}
model by a multiplicative absorption model, namely \texttt{gabs}, at the energy
where such a simple absorption line would be expected for the best-fit magnetic
field strength, that is, $E_\mathrm{gabs} = B_\mathrm{\texttt{cyclofs}} \times
11.57 \,\mathrm{keV} \approx 41.96^{+0.35}_{-0.36}\,\mathrm{keV}$ in the frame of the neutron
star (i.e., before redshifting)
\footnote{Note that the physical cyclotron line model has been
  substituted in place, that is, the \texttt{gabs} model is still within the
  redshift model and therefore shifted to $\sim32.3\,\mathrm{keV}$
  by \texttt{zashift}.}.
Leaving this energy and all other parameters frozen while fitting the width and
the depths of the empirical absorption line results in a reduced $\chi^2$ of
1.57 for 876 degrees of freedom. The corresponding residuals are shown in
Fig.~\ref{fig:fit}c, which clearly illustrates that the centroid energy of the
Gaussian absorption line model is off the cyclotron line. Only when all parameters of the
\texttt{gabs} model are left free, can the profile of the cyclotron line be represented
by a Gaussian absorption line with a centroid energy of $40.01_{-0.14}^{+0.15}\,\mathrm{keV}$ before
redshift (red. $\chi^2$ of 1.17 for 875 d.o.f.).
The value of $E = 40.01\,\mathrm{keV} / (1 + z) = 30.78_{-0.11}^{+0.12}
\,\mathrm{keV}$ after redshift reduction is almost consistent with the value
found by \citet{fuerst15} for the higher energetic Gaussian absorption line used to model
the asymmetric line shape. The ratio of this model to the best-fit model using
the physical CRSF model is shown in Fig.~\ref{fig:fit}d. Evidently, the shapes
of the line models differ significantly as expected.

The model for \object{Cep X-4} presented here makes no claim of
uniqueness. Instead many assumptions are made: other geometries might
fit the spectrum equally well, neglecting bulk velocity becomes
questionable if either the continuum or the cyclotron line are formed
in a region of the accretion column with a significant velocity, and
the usage of an empirical continuum with a ``10 keV'' feature ---
the origin of which is unclear --- is questionable \textsl{per se},
to name just a few. Furthermore, only one viewing angle to the
magnetic field axis is taken into account, which is unrealistic considering
that the data are averaged over pulse phase. The \texttt{cyclofs} model
provides the possibility to average over multiple angles in order to
overcome this inaccuracy albeit in an approximative way (see Sect.~\ref{sec:model_mu}
for details). The width of the cyclotron line is strongly affected by
the viewing angle and the temperature. Here, the width is mainly fitted
by the angle to the magnetic field axis. Magnetic field, temperature,
and velocity gradients are neglected, though they might largely influence
the CRSF line width as well. Studies with more complex configurations
of the CRSF medium are needed for estimating their influence quantitatively.
Physical continuum models should be combined with the physical model for the
CRSF line shape presented here. Their combined application to many observations
of diverse sources covering a large parameter space of both continuum and line
profile model parameters --- some of these parameters might be tied together
--- might help to further constrain the highly degenerate parameter regime.

\section{Summary}
We have described our Monte Carlo code for the generation of synthetic
cyclotron lines. The simulated line profiles have complex shapes and show a
strong dependency on the viewing angle. We have compared our results to the
work from \citet{isenberg98} and find an overall good agreement for both, the
\mbox{slab 1-0} and the \mbox{slab 1-1} geometry.

In order to show the influence of the continuum, the \object{Her X-1} model
spectrum from \citet{becker07} has been used to generate synthetic spectra for
several optical depths and angles to the magnetic field.

A new XSPEC fitting model, \texttt{cyclofs}, which works on precalculated
tables storing the response of the Monte Carlo simulation to monoenergetic
photon injections, has been introduced and is available online (see link in
footnote, page 1).

Using this model, which is describing the cyclotron line shape on physical
grounds, we successfully fitted \mission{NuSTAR} spectra of \object{Cep X-4}.
The resulting magnetic field strength, $B = 3.6 \times 10^{12}\,\mathrm{G}$,
has been found to differ significantly from fits with a Gaussian absorption
line. The reason for this difference might lie in the theoretically complex
profile \citep{isenberg98,schoenherr07,nishimura08} of the fundamental
cyclotron line. This might lead to a different best fit value for the magnetic
field strength --- even for the almost symmetrical and smooth line shapes seen
in observations --- if the modeled CRSF shape is taken into account. Further
studies of more complex physical setups, including the exploration of other
geometries and parameter gradients, the inclusion of angular mixing due to
relativistic light bending, and a combination with an equally physical
continuum model are necessary in order to obtain a fully self consistent
spectrum of the accretion column. The simulation code presented here provides
the flexibility to address these challenges. The associated XSPEC model
mechanism allows for making the corresponding results available in an easily
usable and familiar way.

Other applications of the new code include the simulation of observable
\citep{ferrigno11} phase lags at the CRSF energy \citep{schoenherr14}, a
combination with models for relativistic light bending to obtain self
consistent pulse profiles of the phase dependent CRSF behavior (Falkner et al.,
2016, in prep.), comparisons to observational data from other sources, and a
study of the dependence of the CRSF profile on the magnetic field geometry.

\begin{acknowledgements}
This work has been partially funded by the Deutsche Forschungsgemeinschaft
under DFG grant number WI 1860/11-1 and by the Deutsches Zentrum f\"ur Luft-
und Raumfahrt under DLR grant numbers 50\,OR\,1113, 50\,OR\,1207, 50\,OR\,1410,
and 50\,OR\,1411. We also acknowledge the Russian Foundation for Basic Research
grant number 14-02-91345. MTW is supported by the Chief of Naval Research and
by the National Aeronautics and Space Administration Astrophysical Data
Analysis Program. We thank the International Space Science Institute in Bern
for inspiring team meetings. The fruitful discussions within the MAGNET
collaboration also had a very positive impact on this work. The figures in this
work have been produced using the \texttt{slxfig} package by \citet{slxfig}.
\end{acknowledgements}

\bibliographystyle{aa}
\tiny{\bibliography{paper}\par}

\appendix

\section{The XSPEC model}\label{sec:model}
Our XSPEC model uses precalculated Green's function tables to imprint
synthetic CRSFs on arbitrary continua. These Green's tables have
been calculated from the response of a medium to monoenergetic photon
injection on a wide range of parameters. Therefore each geometry has
its own Green's function table, which can be selected for use by
\begin{itemize}
\item setting the environment variable \textit{CYCLOFS\_TABLE} to the
  table's location
\item defining the table location as initialization string in the
  \texttt{model.dat} file
\end{itemize}

Table~\ref{table:model_pars} lists all model parameters. At least one
optical depth, the magnetic field, the temperature, and the angle or
the angular range and the number of angular points must be set for a
successful model evaluation.

\begin{table}
  \caption{XSPEC model parameters.}\label{table:model_pars}
  \centering
  \begin{tabular}{cp{0.7\columnwidth}}
\hline\hline
Name & Description \\
\hline
tau\_para & Parallel optical depth [$\tau_{\mathrm{T}}$]      \\
tau\_perp & Perpendicular optical depth [$\tau_{\mathrm{T}}$] \\
B         & Magnetic field [$10^{12}\,$G]                                    \\
T         & Electron temperature $k_\mathrm{B}T$ [keV]                              \\
mu        & Cosine of the viewing angle to the $B$-field, $\mu = \cos\vartheta$ \\
mu\_N     & Number of mu-values to average over (0 for exact value of mu)           \\
mu\_min   & Minimum value of mu for averaging                                       \\
mu\_max   & Maximum value of mu for averaging                                       \\
\hline
  \end{tabular}
\end{table}

\subsection{Interpolation}\label{sec:model_interp}
The model can be configured to interpolate between parameter values in different ways depending on the
parameter: currently a linear interpolation scheme is utilized for all parameters.
Model spectra are also extrapolated, if the desired parameter
combination is not covered by the Green's table. A linear scheme is used
here, as well.
The model prints out a warning message if
extrapolation is used. The extrapolation method used for the optical depth is
slightly different from the others: For a given optical depth out of the table
range, the model convolves its output iteratively with a suitable optical depth
within the table range. This extrapolation-via-successive-convolution method
turned out to be as accurate as linear extrapolation close to the boundaries
but yields much better results than any other method for extrapolation over
orders of magnitude. This is especially useful for studies in the regime of
high optical depths where calculation time increases significantly. The
accuracy of such an extrapolation over orders of magnitude is questionable,
though, as it depends on the assumption of isotropic angular redistribution,
which is normally not justified, as we have shown before \citep{schoenherr14}.
It is, nevertheless, very useful for studying the influence of increased line
depth on the overall combined model flux and therefore included as default
behavior with a warning message.

\subsection{Angular averaging}\label{sec:model_mu}
The model is designed to calculate spectra for exactly one viewing angle,
defined by its cosine \textit{mu}$ = \cos\vartheta$, to the
magnetic field axis. It provides the possibility for averaging over a range
of angles to the $B$-field axis, by returning the mean value of
\textit{mu\_N} spectra between \textit{mu\_min} and \textit{mu\_max} for each
energy bin. If \textit{mu\_N} is set to 0, only one spectrum for the angle
\textit{mu} is calculated. If it is set 1, one spectrum right in the middle of
the angular range specified by \textit{mu\_min} and \textit{mu\_max}, that is,
for an angle $\mu = 0.5 (\mu_\mathrm{min} + \mu_\mathrm{max})$, is returned. A
\textit{mu\_N} value of 2 will result in a spectrum averaged from two points,
namely \textit{mu\_min} and \textit{mu\_max}. For higher values, the additional
points are equally spread over the angular range. Note that the parameter
\textit{mu} is not used at all if \textit{mu\_N} is larger than zero and should
be frozen during fitting in order to avoid useless iterations. The same applies
to the parameters \textit{mu\_min} and \textit{mu\_max} in the case of
$\mathrm{\textit{mu\_N}} = 0$.

\end{document}